\documentstyle[pra,multicol,aps]{revtex}
\begin{document}

\input{epsf.tex}
\epsfverbosetrue
\title{Nonlinear Theory of Soliton-Induced Waveguides}

\author{Elena A. Ostrovskaya and Yuri S. Kivshar}

\address{Optical Sciences Centre, Research School of
         Physical Sciences and Engineering\\
         Australian National University, Canberra
         ACT 0200, Australia}
\maketitle
\begin{abstract}
We develop a nonlinear theory of soliton-induced waveguides
that describes a finite-amplitude probe beam guided by a spatial dark soliton, in a saturable nonlinear medium. We suggest an effective way of controlling the interaction of these soliton-induced waveguides and also show
that, in sharp contrast with scalar dark solitons, the
dark-soliton waveguides can attract each other and even
form stationary bound states.
\end{abstract}

\begin{multicols}{2}
\narrowtext

It was suggested in the early days of nonlinear optics, that the waveguiding properties of a self-trapped laser beam could be used to guide atoms or ions along the beam's axis \cite{ask}. A similar
physical mechanism is responsible for the trapping of a weak beam by a spatial (bright or dark) optical soliton. This
phenomenon is usually referred to as the soliton-induced
waveguiding of a probe beam. It has been widely discussed
theoretically and has also been observed experimentally
\cite{snyder,manas,barth,barry,moti}.

As has already been established, the major advantage of
employing the waveguiding properties of dark, rather than
bright, spatial solitons is their greater stability and
steerability in a Kerr medium \cite{dark_review}. Moreover, in
the case of saturable nonlinearity, dark-soliton waveguides remain single-moded \cite{note}, unlike the waveguides created by bright solitons \cite{moti}.

The theory of soliton-induced waveguides developed so far
is based on the assumption that a probe beam guided by
a spatial soliton is {\em weak} (see, e.g., Refs.
\cite{manas,barth,moti}). However, when the intensity of the
probe beam is increased, the guided mode requires a broader
waveguide and the beam interacts strongly with the waveguide,
changing it in a self-consistent manner. Such a structure can
not be described by a linear waveguide theory because it represents a
bound state of the coupled bright and dark components. In this
Letter we develop a nonlinear theory of the soliton-induced
waveguides, treating the case of two incoherently interacting
beams in a photorefractive medium as a model already realized
experimentally \cite{bright_dark}. We also analyze the
interaction of two neighboring waveguides and show that
attractive forces acting between the guided beams can result
in suppression of the repulsion between dark solitons.

Following the original paper by Christodoulides {\em et al.}
\cite{segev}, we consider two incoherently interacting linearly
polarized beams with the normalized envelopes $U$ and $W$ in a
photorefractive medium, described by the equations,
\begin{equation} \label{eq_UW} \begin{array}{l} {\displaystyle
i \frac{\partial U}{\partial z} + \frac{1}{2} \frac{\partial^2
U}{\partial x^2} + \frac{ \beta (1 + \rho) U}{ 1 + |U|^2 +
|W|^2}  \hspace{2.4mm} =  0,} \\*[9pt] {\displaystyle i
\frac{\partial W}{\partial z} + \frac{1}{2} \frac{\partial^2
W}{\partial x^2} + \frac{ \beta (1 + \rho) W}{ 1 + |U|^2 +
|W|^2}  \hspace{2.4mm} = 0.}
\end{array}
\end{equation}
Here $\rho = I_{\infty}/I_d$, $\beta = (k_0x_0)^2 n_e^4
r_{33} E_0/2$, where $I_{\infty}$ stands for the total power density
away from the beam, $I_d$ is the so-called dark irradiance,
$k_0$ is the propagation constant, $x_0$ is the spatial width
of the beam, and $n_e^4 r_{33}E_0$ is a correction to the
refractive index that it due to the external field applied to a crystal
along the $x$-axis \cite{segev}.

Localized solutions of Eqs. (\ref{eq_UW}) can be sought in
the form $U= \sqrt{1-q_1} \, u \, e^{iq_1z}$ and $W =
\sqrt{1-q_1} \, w \,  e^{iq_2 z}$, and they form a
continuous two-parameter family.  System (\ref{eq_UW}) can be
further simplified by measuring the spatial coordinates $z$
and $x$ in units of $\sqrt{s}/\beta (1 + \rho)$ and
$\{s/[\beta(1+\rho)]\}^{1/2}$, respectively, where $s= 1-q_1$.
Then, $u$ and $w$ satisfy the normalized equations:
\begin{equation}
\label{eq_uw}
\begin{array}{l} {\displaystyle
i \frac{\partial u}{\partial z} + \frac{1}{2} \frac{\partial^2
u}{\partial x^2} - \frac{(|u|^2 + |w|^2)u}{ 1 + s(|u|^2
+|w|^2)} + u \hspace{2.4mm} =  0,} \\*[9pt] {\displaystyle i
\frac{\partial w}{\partial z} + \frac{1}{2} \frac{\partial^2
w}{\partial x^2} - \frac{(|u|^2 + |w|^2)w}{ 1 + s(|u|^2 +
|w|^2)} + \lambda w \hspace{2.4mm} = 0,} \end{array}
\end{equation} where $\lambda = (1-q_2)/s < 1$ is the
dimensionless soliton parameter, and $s$ characterizes
{\it nonlinearity saturation}, $(s<1)$. In the limit $s \rightarrow 0$,
Eqs. (\ref{eq_uw}) reduce to the Manakov model which possesses
exact bright and dark multisoliton solutions \cite{manakov1,nogami,nlgw_sk,manakov}.

Stationary solutions are described by Eqs. (\ref{eq_uw})
with $z$ derivatives omitted. We assume that the component
$u$ has nonvanishing asymptotics, whereas $w$ is spatially localized.
Then the simplest solution of this kind describes a dark
soliton of the field $u$, provided that $w \equiv 0$. This solution
can be presented in quadratures and then found numerically
\cite{moti}. For the component $w$, the dark soliton creates an
effective waveguide. To analyze what kind of guided mode can
be supported by this waveguide we linearize the equation for
$w$ and study its localized solutions. That is, we consider a
one-component dark soliton $u=u_s(x)$ and add a small
perturbation written as $w = {\rm O}(\epsilon)$ and $u = u_s(x)
+ {\rm O}(\epsilon^2)$.  It is easy to verify from Eqs.
(\ref{eq_uw}) that O($\epsilon$) in the expansion for
$u$ is zero.  Substituting this expansion into Eqs.
(\ref{eq_uw}), we obtain two decoupled equations, of which the
stationary equation for $w$ is \begin{equation}
\label{eigen}
\frac{1}{2} \frac{d^2 w}{d x^2} - \frac{|u_s|^2}{1 + s|u_s|^2}
w  + \lambda w = 0.  \end{equation}
Depending on the properties of the effective
 waveguide $V(x) \equiv |u_s(x)|^2 (1 + s |u_s(x)|^2)^{-1}$,
eigenvalue problem (\ref{eigen}) possesses a number of solutions that
decay as $x \rightarrow \pm \infty$, the so-called {\em guided}
or {\em bound} modes. In the limit $s \rightarrow 0$, the symmetric
solution $w_1(x) = {\rm sech} \, x$ exists for any $\lambda
> \lambda_0 = 1/2$. In general, the cutoff of a bound mode,
$\lambda_0(s)$, depends on the saturation parameter $s$.
Furthermore, if $u_s(x)$ is a dark soliton of a saturable
medium, there exists {\em only one bound mode} of eigenvalue problem (\ref{eigen}) for any value of $s<1$, as
was also found in Ref. \cite{moti}. Therefore the dark-soliton
waveguide in a saturable medium is always single moded \cite{note}.
\begin{figure}
\setlength{\epsfxsize}{7.0cm}
\centerline{\epsfbox{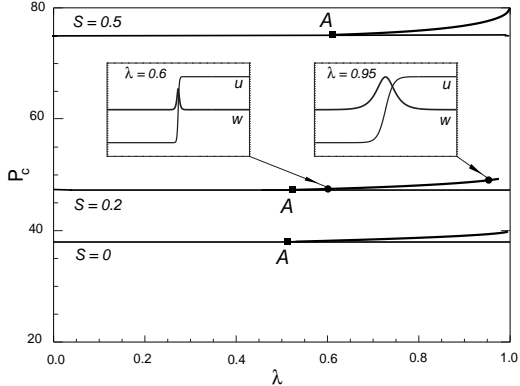}}
\vspace{0.2cm}
\caption{Bifurcation of a one-component dark soliton $u_s(x)$
for three values of $s$, shown as
the dependence of the complimentary power $P_c$ on the
propagation constant $\lambda$. Examples of two-component localized solutions are shown for $s=0.2$.}
\vspace{-0.4cm}
\label{fig1}
\end{figure}
\begin{figure}
\setlength{\epsfxsize}{6.0cm}
\centerline{\epsfbox{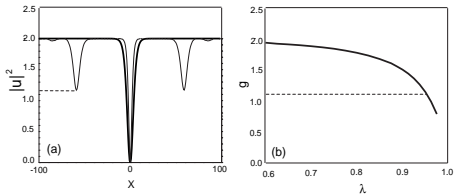}}
\vspace{0.2cm}
\caption{Evolution of the dark-soliton waveguide without
the guided mode for $s = 0.5$. (a) Soliton profiles at $z=0$ (thick line) and $z=
100$ (thin line) for $\lambda = 0.95$. (b) Amplitude of the additional
gray-soliton pair $g$ versus $\lambda$ measured at $z=
100$.}
\vspace{-0.4cm}
\label{fig2}
\end{figure}
\begin{figure}
\setlength{\epsfxsize}{7.5cm}
\centerline{\epsfbox{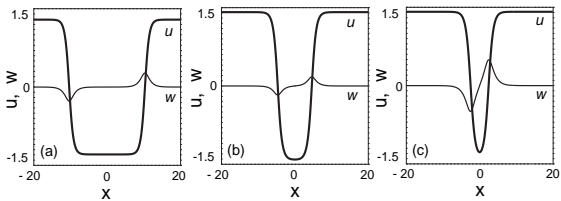}}
\vspace{0.2cm}
\caption{Examples of two-soliton bound states in model
(\ref{eq_uw}): (a) $s = 0.4$ and $\lambda = 0.55$, (b) $s= 0.5$
and $\lambda = 0.55$, (c) $s= 0.5$
and $\lambda = 0.65$.}
\vspace{-0.4cm}
\label{fig3}
\end{figure}
The localized solution that describes the fundamental (symmetric)
mode of a dark-soliton waveguide appears as a bifurcation of
the one\--component dark  soliton, $u_s$, in which a
two\--component solution $(u,w)$ emerges from the
one\--component solution $(u_s,0)$ at certain $\lambda =
\lambda_0(s)$. Such localized solutions form a continuous
family in  $\lambda$ for a fixed $s$.  Figure 1 presents
examples of such a bi\-fur\-ca\-tion together with the corresponding two-component soliton states, found numerically
by the relaxation technique, for different values of $s$.  Shown is the
dependence of the complimentary power, \[ P_c =
\int_{-\infty}^{+\infty} dx \left\{|u(x)|^2 + |w(x)|^2 -
(1-s)^{-1}\right\}, \] on the soliton parameter, $\lambda$.

From the physical point of view, the parameter $\lambda$
characterizes the intensity of a bright component guided by
a dark-soliton waveguide. As $\lambda$ grows, the intensity of the bright component grows as well, remaining less than that of the dark-soliton background. As it approaches a maximum, the dark soliton becomes broader. This
means that, for $\lambda$ much larger than $\lambda_0(s)$, the soliton waveguide deforms: It becomes wider to guide a beam of higher intensity. This effect is definitely beyond the linear approximation provided by the model [Eq. (\ref{eigen})]. It can, however, be described by a nonlinear theory constructed by numerical solution of coupled equations (\ref{eq_uw}). The corresponding
solutions, presented in Fig. 1 for $\lambda > \lambda_0(s)$,
generalize the localized solutions of the linear eigenvalue problem
[Eq. (\ref{eigen})] valid only near the bifurcation points $A$ where $\lambda \sim \lambda_0$.

To verify the concept of a deformed waveguide, we performed
simulations of the waveguide propagation in the case when the
bright component is removed. As expected from the theory of
dark solitons \cite{dark_review} and previously confirmed in \cite{carvalho}, a wider dip in the input beam
generates at least one additional pair of gray solitons, as is shown in
Fig. 2(a). The amplitude of the additional gray-soliton pair $g$
is shown in Fig. 2(b) as function of the parameter $\lambda$.

Similar to the case of the exactly integrable Manakov system
\cite{nlgw_sk,manakov}, the nonintegrable model (\ref{eq_uw}) supports stationary bound states of two (or more) solitons. We have found such stationary structures numerically. Examples of the stationary solutions, corresponding to different values of $\lambda$ and $s$, are shown in Figs. 3 (a,b,c).

 The stationary bound states are expected when the total
force acting between the neighboring two-component solitons
vanishes. Two closely separated dark
solitons always repel each other, and this property does not
depend on integrability \cite{dark_review}. In the light of
this knowledge, the existence of stationary solutions in the
form of two bound dark-soliton waveguides is rather surprising.
Therefore, we are led to wonder whether the introduction of a
bright component causes {\em an attractive force} between
solitons that could nullify the repulsion of the dark
components. Note that the stationary bound states of the model
[Eqs. (\ref{eq_uw})] are formed by bright components that are {\em out of
phase} with each other (Figs. 3). Expecting
them to attract each other would be inconsistent with the theory of
bright solitons that predicts attraction only between
in-phase solitons. However, in our case the bright component
by itself is not a soliton because, after the dark component is removed, it
diffracts in the defocusing medium. Localization of the
bright components is due to trapping in an effective waveguide created by dark components. As a result, interaction of these guided modes is
different to that expected for bright solitons. From the physical point of view, even slight overlap between two out of phase localized beams in a defocusing medium leads to an increase of the refractive index in the overlapping region, therefore increasing the attraction of the beams.
\begin{figure}
\setlength{\epsfxsize}{6.0cm}
\centerline{\epsfbox{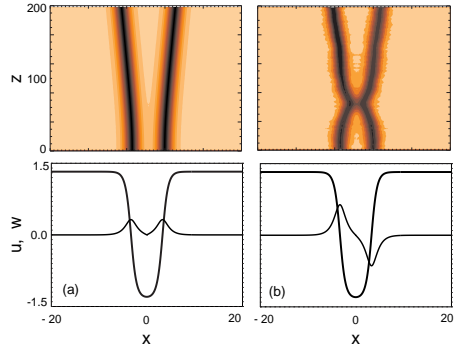}}
\vspace{0.2cm}
\caption{Interaction between two dark-soliton
waveguides guiding bright components (a) in phase or
(b) out of phase.}
\vspace{-0.4cm}
\label{fig4}
\end{figure}
Analyzing interaction between the soliton waveguides, we look
for a solution to describe two weakly overlapping
two-component solitons in the form:  $u = \tanh (x-x_0) \,
\tanh (x+x_0)$ and $w = a \, {\rm sech} (x-x_0) + a \,
e^{i\phi} {\rm sech} (x+x_0)$.  Then we employ the Lagrangian
formalism for dark solitons \cite{krolik} to derive effective
equations for the soliton parameters, including the soliton
separation $2x_0$ and the relative phase between the bright
components $\phi$. Omitting lengthy details, we present the
effective energy of the soliton interaction in the form $(x_0 \gg 1)$,\begin{equation} \label{poten} 
U_{\rm eff}(x_0) =  2 e^{-4x_0} + 4 a^2 \, x_0 e^{-2 x_0}\cos (\phi).  
\end{equation} The first
term in Eq. (\ref{poten}) describes the interaction of
two scalar dark solitons; see Eq. (34) in Ref.  \cite{krolik}.
 The second term appears as the result of the interaction between the
 bright components of the amplitude $a$. As follows from Eq.
(\ref{poten}), the repulsive force between dark components is
{\em weaker} than the force acting between the bright
components. Moreover, introducing bright components into the
closely separated dark solitons can lead to a nontrivial effect
when the repulsion of the dark components is compensated for by an
attractive force acting between the bright components.  This effect
requires that the bright components be out of phase
$(\phi =\pi)$, in agreement with the stationary solutions found
numerically; see Fig. 3. Figure 4
shows the results of the propagation of two neighboring
dark-soliton waveguides when the bound modes they guide are in
phase [Fig.  4(a)] or out of phase [Fig 4(b)]. In the former
case the interaction between bright components is repulsive,
and the solitons repel each other even more strongly than two scalar dark
solitons. In the latter case, the interaction is attractive and
it forces the dark solitons to collide. These results confirm that the
interaction of the dark-soliton waveguides is phase
sensitive. We believe that the demonstrated behaviour can be observed experimentally for the incoherent interaction of photorefractive solitons.

In conclusion, we have developed a nonlinear theory of the
soliton-induced waveguides created by dark solitons in
a saturable optical medium. We have found families of the soliton states
describing a finite-amplitude guided mode trapped by a dark
soliton. We have described a novel type of interaction of the
dark-soliton waveguides in which the repulsion of the neighboring
dark solitons is suppressed by an attractive force acting
between the guided modes, leading to the existence of
stationary two-soliton bound states.

This research has been supported by the Australian Photonics
Cooperative Research Centre. The authors are indebted to
D. Christodoulides, M. Segev, A. Sheppard, and A. Snyder
for useful discussions.

\end{multicols}
\end{document}